\documentclass[preprint,5p,sort&compress]{elsarticle}

\makeatletter
\def\ps@pprintTitle{%
 \let\@oddhead\@empty
 \let\@evenhead\@empty
 \def\@oddfoot{}%
 \let\@evenfoot\@oddfoot}
\makeatother

\usepackage{graphicx,amssymb,amsmath,amsthm,amsfonts,epsfig,epsf}
\usepackage[linktocpage]{hyperref}
\usepackage[usenames, dvipsnames]{color}
\usepackage{epstopdf}

\usepackage{bm}
\usepackage{dcolumn}
\usepackage[latin1]{inputenc}
\usepackage{latexsym}
\usepackage{rotating}
\usepackage{hyperref}
\usepackage{longtable}
\usepackage{enumerate}
\usepackage{tensor}
\usepackage{url}
\usepackage[normalem]{ulem}
\renewcommand{\vec}[1]{\boldsymbol{#1}}

\def\be{\begin{equation}}
\def\ee{\end{equation}}
\newcommand{\beq}{\begin{eqnarray}}
\newcommand{\eeq}{\end{eqnarray}} 
\newcommand{\ba}{\begin{align}}
\newcommand{\ea}{\end{align}}
\newcommand{\bsub}{\begin{subequations} \begin{eqnarray}}
\newcommand{\esub}{\end{eqnarray} \end{subequations}}
\newcommand{\p}{\partial}

\newcommand{\om}{\omega}

\newcommand{\tom}{\widetilde \om}

\begin{document}

\begin{frontmatter}

\title{Synchronized stationary clouds in a static fluid}

\author[ad1]{Carolina L.~Benone}
\ead{lben.carol@gmail.com}

\author[ad1]{Lu\'is C.~B.~Crispino}
\ead{crispino@ufpa.br}

\author[ad3,ad5]{Carlos A.~R.~Herdeiro}
\ead{herdeiro@ua.pt}

\author[ad4]{Maur\'icio Richartz}
\ead{mauricio.richartz@ufabc.edu.br}


\address[ad1]{Faculdade de F\'isica, Universidade Federal do Par\'a, 66075-110, Bel\'em, Par\'a, Brazil}
\address[ad3]{Departamento de F\'isica da Universidade de Aveiro and Center for Research and Development in Mathematics and Applications (CIDMA), Campus de Santiago, 3810-183 Aveiro, Portugal}
\address[ad5]{Centro de Astrof\'\i sica e Gravita\c c\~ao - CENTRA, Departamento de F\'\i sica,
Instituto Superior T\'ecnico - IST, Universidade de Lisboa - UL, Avenida
Rovisco Pais 1, 1049-001, Portugal}
\address[ad4]{Centro de Matem\'atica, Computa\c{c}\~ao e Cogni\c{c}\~ao, Universidade Federal do ABC (UFABC), 09210-170 Santo Andr\'e, S\~ao Paulo, Brazil}

\begin{abstract}
The existence of stationary bound states for the hydrodynamic velocity field between two concentric cylinders is established. We argue that rotational motion, together with a trapping mechanism for the associated field, is sufficient to mitigate energy dissipation between the cylinders, thus allowing the existence of infinitely long lived modes, which we dub \textit{stationary clouds}. We demonstrate the existence of such stationary clouds for sound and surface waves when the fluid is static and the internal cylinder rotates with constant angular velocity $\Omega$. These setups provide a unique opportunity for the first experimental observation of \textit{synchronized} stationary clouds. As in the case of bosonic fields around rotating black holes and black hole analogues, the existence of these clouds relies on a synchronization condition between $\Omega$ and the angular phase velocity of the cloud.
\end{abstract}


\end{frontmatter}

\section{Introduction}

With a few notable exceptions, like in superconductivity, dissipation plays an important and decisive role in Physics. Due to dissipation, the oscillations of any perturbed system (for instance a ringing bell) will die away with the passing of time. In general, waves propagating in realistic (hence dissipative) media will lose energy and hence decrease their amplitude in time. Black hole (BH) absorption in a scattering process and BH response to perturbations (quasinormal modes), for example, can be interpreted as manifestations of dissipation: the event horizon acts as a one-way membrane which extracts energy from the BH exterior, dissipating any external perturbation~\cite{review1,review2,Berti:2009kk,Konoplya:2011qq}.

Linear perturbations of a dissipative system will generically decay in time according to $\exp\left(-i \omega t \right)$, where $\omega$ is a complex frequency. Combined with rotational motion, however, dissipation can be mitigated and unexpected phenomena arise. One such example is superradiance, a scattering process in which low-frequency modes are amplified by a rotating object~\cite{zeldovich1,bek_1,Richartz:2009mi,Brito:2015oca}. This amplification effect can be understood as the dissipation of negative energies (with respect to the rest frame of the rotating object), meaning that low-frequency waves can lower the energy of the rotating system. In other words, while in the laboratory frame the flux of dissipated energy is proportional to the oscillation frequency $\omega$ of the waves, in the rotating frame the energy flux is proportional to the effective frequency $\tilde \omega = \omega - m \Omega$, where $m$ is the azimuthal number of the wave and $\Omega$ is the angular velocity of the scatterer. Therefore, sufficiently low-frequency waves satisfying $\tilde \omega < 0$ will reverse the energy flux direction, thus extracting energy from the rotating object.

If, besides rotational motion, an additional trapping mechanism is present in the system, a feedback process, in which successive superradiant amplifications occur, is triggered. As a result, the system will exhibit superradiant instabilities which, at the linear level, grow exponentially in time. Such unstable modes have $\mathrm{Im}(\omega) > 0$. At the threshold between stable and unstable modes, \textit{stationary} bound states, characterized by $\mathrm{Im}(\omega) = 0$, exist. Such bound states, dubbed \textit{stationary clouds}, are  generic features of rotating, dissipative systems. Nevertheless, they have never been experimentally observed.

Stationary clouds have played a key role, for instance, in recent developments in BHs physics. They appear when a scalar wave is synchronized with a rotating BH, meaning that the wave's angular phase velocity, $\omega/m$, matches the angular velocity $\Omega$ of the BH horizon. These stationary bound states are infinitely long lived massive scalar field configurations which decay exponentially at asymptotic infinity, and that exist under a certain quantization condition. Such solutions were first identified by Hod~\cite{Hod:2012px}, and further investigated in, $e.g.$, Refs.~\cite{Hod:2014baa,Hod:2013zza,benone2014kerr,Benone:2014nla,Hod:2014pza,Hod:2014sha,Herdeiro:2014pka,Wilson-Gerow:2015esa,Wang:2015fgp,0264-9381-33-15-154001,Bernard:2016wqo,Sakalli:2016xoa,Hod:2016lgi,Hod:2016yxg,Herdeiro:2017oyt,Ferreira:2017cta}.
Going beyond the linear approximation, Herdeiro and Radu~\cite{Herdeiro:2014goa} (see also Refs.~\cite{Herdeiro:2015gia,Chodosh:2015oma}) have shown that rotating BHs with scalar hair exist as fully non-linear solutions of the coupled Einstein-Klein-Gordon equations, being the non-linear realization of rotating BHs surrounded by (test field) stationary scalar clouds -- see Ref.~\cite{0264-9381-33-15-154001} for the analogous case of Kerr BHs with Proca hair, Refs.~\cite{East:2017ovw,Herdeiro:2017phl} for a discussion of their formation properties and Refs.~\cite{Ganchev:2017uuo,Degollado:2018ypf} for a discussion of their stability.
These BHs have provided an influential counter-example to the no-hair conjecture in General Relativity with matter that does not violate energy conditions -- see Ref.~\cite{Herdeiro:2015waa} for a review.
In this scenario, the trapping mechanism is provided by the field's mass.
 
Instead of a mass term, another way to confine waves and allow for stationary bound states is to have a reflective boundary at a sufficient large distance from the BH - see, $e.g.$, Refs.~\cite{Hod:2014pza,Benone:2014nla}. For astrophysical BHs it is hard to imagine a realistic way to accomplish this. Nonetheless, analogue BHs~\cite{Unruh:1980cg,Barcelo:2005fc}, being reproducible in the laboratory, can be  encompassed by a reflective surface, as discussed in Ref.~\cite{Benone:2014nla} (see also Ref.~\cite{PhysRevLett.121.061101}, which shows that non-zero vorticity can act as an effective mass term, thus allowing quasibound states to appear in vortex flows). For superradiance, the presence of an event horizon is not mandatory~\cite{Richartz:2009mi,Richartz:2013hza,Oliveira:2014oja,Oliveira:2018ckz}: it can be replaced by any dissipative material, as in the case of Zel'dovich's cylinder that scatters electromagnetic waves~\cite{zeldovich2} or in the case of a rotating cylinder that dissipates the energy of  water waves~\cite{cylinder}. Here,  building on the ideas of Refs.~\cite{Benone:2014nla,cylinder} we analyze the possibility of using water waves to set up a laboratory experiment to observe, for the first time, synchronized stationary clouds.

 \section{Sound waves and surface waves between two concentric cylinders}

The Analogue Gravity programme for fluids, initiated by the seminal work of Unruh~\cite{Unruh:1980cg}, is based on the fact that sound waves in a background flow can mimick the propagation of scalar fields in General Relativity. It relies on the assumptions of barotropicity, inviscidity, and irrotationality to establish that velocity perturbations of a fluid flow satisfy the Klein-Gordon equation for scalar fields in a curved spacetime. Important gravitational effects like Hawking radiation and superradiance can be experimentally tested within such analogies~\cite{stim_HR,germain_watertank,steinhauer,SR_obsvn}.  

Here we are primarily interested in the setup proposed in~\cite{cylinder}. It consists of a rotating cylinder, with angular speed $\Omega$ and radius $R_0$, surrounded by an external static cylinder of radius $R_1>R_0$. The cylinders are concentric and the space between them is filled with a static fluid. Both cylinders will generally dissipate the energy that they absorb from waves impinging on them.
The details of such dissipation are encoded in the \textit{impedances} $Z_\om^{\mathrm{in}}$ and $Z_\om^{\mathrm{out}}$ of the inner and outer cylinders, respectively. 

Impedances are complex quantities~\cite{Rienstra,ginsberg} depending on various factors like wave frequency,  geometry of the absorber and material properties, that determine the boundary conditions at the surface of the cylinders. According to Ref.~\cite{Rienstra}, impedance is ``a measure of the lumpiness of the surface". Mathematically, it is defined as the ratio between the complex amplitude of pressure at the boundary and the complex amplitude of particle velocity into the boundary~\cite{ginsberg}.
The real part of the impedance is called \textit{resistance} and relates to the non-conservative part of the energy flow. The imaginary part, called \textit{reactance}, cyclically stores and releases energy without dissipation~\cite{kennelly,Rienstra}.

 The role of acoustic impedance is understood through an analogy with a mass-damper-spring system~\cite{ginsberg}: while the damping coefficient corresponds to $\mathrm{Re}(Z_\om)$, the mass and spring terms both contribute to $\mathrm{Im}(Z_\om)$. If the mass term dominates, we have $\mathrm{Im}(Z_\om)<0$, characterizing a stiffer surface (inertance regime). If, on the other hand, the spring term dominates, we have $\mathrm{Im}(Z_\om)>0$, characterizing a more compliant surface (compliance regime).  Passive surfaces, as opposed to active surfaces, do not produce energy and are characterized by $\mathrm{Re}(Z_\om) \leqslant 0$.
 Below we will show that irrespectively of what the impedance of the inner cylinder is (i.e.~active/passive and inert/compliant), by synchronizing wave motion with the cylinder's rotation we can halt the flux of energy through its surface. The impedance of the outer cylinder is, nevertheless, important. We shall be interested in the case of negligible energy dissipation at the external surface, corresponding effectively to $\mathrm{Re}(Z_\om) = 0$.
Then, the wave energy is completely reflected and $\mathrm{Im}(Z_\om)$ controls the phase difference between the ingoing and outcoming waves. For instance, in the case of ``hard walls", characterized by $\mathrm{Re}(Z_\om) \rightarrow 0$ and $|\mathrm{Im}(Z_\om)| \rightarrow \infty$, this phase difference will be of $\pi$, corresponding to a Neumann boundary condition for the wave field. ``Soft walls", on the other hand, characterized by $\left|Z_\om\right| \rightarrow 0$,
 do not generate phase differences, corresponding to a Dirichlet boundary condition for the wave field.

Velocity perturbations $\delta \vec{v}$ of the background static flow $\vec{v}=0$ can be described in terms of a scalar field $\psi$ by the relation $\delta \vec{v} = \nabla \psi$. From the hydrodynamical equations, one can  show that wave motion is determined by the Klein--Gordon equation
\be 
\nabla_\mu \nabla^{\mu} \delta \psi = -\frac{1}{\sqrt{-g}}\frac{\partial}{\partial x^{\mu}} \left(g^{\mu \nu} \sqrt{-g} \frac{\partial \delta \psi}{\partial x^{\nu}} \right) = 0,
\ee   
where $x^{\mu}=(t,r,\theta,z)$ is the cylindrical coordinate system in the laboratory frame, $g_{\mu \nu} = \eta \cdot \mathrm{diag}\left(-c^2,1,r^2,1\right)$ is the analogue metric, and $g=\mathrm{det}(g_{\mu \nu})$. Here, $c$ denotes the sound velocity (in the case of sound waves) or the gravity wave speed (in the case of shallow water surface waves). For sound waves, the constant $\eta$ is given by $\rho/c$, where $\rho$ is the fluid density; for surface waves, $\eta$ equals $h/c$, where $h$ is the depth of the fluid.

Due to the cylindrical symmetry of the setup, the wave equation can be reduced to an ordinary differential equation for the radial coordinate $r$. Indeed, with the \textit{ansatz}
\beq 
\label{modesum}
\psi(t,r,\theta, z) = \frac{\varphi(r)}{\sqrt{r}}  e^{-i\omega t+i m \theta },
\eeq
the Klein-Gordon equation reduces to
\be 
\p_r^2 \varphi + \left[\frac{\omega^2}{c^2} - \frac{1}{r^2}\left(m^2 - \frac{1}{4}\right) \right] \varphi = 0  \label{eq:radial}, 
\ee 
where $\omega$ is the frequency of the wave and $m \in \mathbb{Z}$ is the azimuthal wave number. Its most general solution is given in terms of Bessel functions, namely
\be \label{wavesol}
\varphi(r) = D_1 \sqrt{r} J_m\left( \frac{\omega r}{c}\right)+ D_2 \sqrt{r} Y_m\left( \frac{\omega r}{c}\right),
\ee
where $D_1$ and $D_2$ are constants.

\section{Boundary conditions for stationary clouds}
Waves described by Eq.~\eqref{wavesol} propagate in the region $R_0<r<R_1$ between the cylinders. 
At the surface of the external cylinder, for both sound and surface waves, the boundary condition is given in terms of the impedance by~\cite{Rienstra,cylinder}:
\be \label{BCcyl0}
\left. \left( \frac{\p_r \psi}{\psi} \right) \right|_{r=R_1} =  - \frac{i \rho \om}{Z_{\om}^{\mathrm{out}}}.
\ee
The boundary condition above, after replacing $R_1$ by $R_0$ and $Z_{\om}^{\mathrm{out}}$ by $Z_{\om}^{\mathrm{in}}$, also holds for the inner cylinder, but only in its rest frame. Since the inner cylinder rotates uniformly with angular velocity $\Omega$, it is necessary to transform the angular coordinate $\theta$ to a new angular coordinate $\widetilde \theta=\theta+\Omega t$ before applying the boundary condition. This is equivalent to replacing $\omega$ with $\tom = \omega-m\Omega$ in \eqref{BCcyl0}, so that  
 \be \label{BCcyl}
\left. \left( \frac{\p_r \psi}{\psi} \right) \right|_{r=R_0} =  - \frac{i \rho \tom}{Z_{\tom}^{\mathrm{in}}}
\ee
in the laboratory frame.

Energy dissipation implies that waves propagating between the cylinders will, in general, decay in time with the characteristic timescale $\tau \sim 1/|Im(\om)|$.
 Due to superradiance, however, some modes (satisfying $0<\mathrm{Re}(\omega)<m\Omega$) will instead grow in time at least until non-linearities become important. Both cases were analyzed in Ref.~\cite{cylinder}. Here, instead, we focus on the possibility of canceling the dissipation effects at the inner cylinder. 

To accomplish this we consider waves that are \textit{synchronized} with the rotating cylinder, so that $\omega/m=\Omega$ holds. Consequently, the term proportional to $\tom$ in the boundary condition \eqref{BCcyl} vanishes,
making the impedance of the inner cylinder irrelevant for the problem and establishing that any dissipative effect will be completely suppressed. In contrast, the authors of Ref.~\cite{cylinder} were interested in superradiance and superradiant instabilities, hence it was crucial for them that the inner cylinder was a passive surface. 
Regarding the external static cylinder, we assume that the energy flow through it is negligible, so that, effectively,
$\mathrm{Re}\left(Z_\omega^{\mathrm out}\right) = 0$.  We investigate the formation of stationary clouds for several pure imaginary values of impedance ranging from the hard wall limit, $|\mathrm{Im}(Z_\om)| \rightarrow \infty$, to the soft wall limit, $|\mathrm{Im}(Z_\om)| \rightarrow 0$.

Combining Eq.~\eqref{wavesol} with the synchronization condition $\omega=m\Omega$, it is straightforward to show that the boundary conditions \eqref{BCcyl0} and \eqref{BCcyl} will be satisfied only when
\be
\frac{J'_m}{Y'_m}=\frac{i\hat{J}_m-Z^{\mathrm{out}} \hat{J}'_m}{i\hat{Y}_m- Z^{\mathrm{out}} \hat{Y}'_m},
\label{mbc2}
\ee
where $Z^{\mathrm{out}}= Z_{m\Omega}^{\mathrm{out}}/(\rho c)$, $J_i=J_i(m\alpha)$, $Y_i=Y_i(m\alpha)$, $\hat{J}_i=J_i(m\alpha R_1/R_0)$, $\hat{Y}_i=Y_i(m\alpha R_1/R_0)$, and
$\alpha = \Omega R_0/c$.
The derivatives, denoted by $'$, are taken with respect to the whole argument of the Bessel functions. Note that the equation above depends only on four parameters: $\alpha$, $m$, $R_1/R_0$, and $Z^{\mathrm{out}}$. For given  values of $m$, $R_1/R_0$, and $Z^{\mathrm{out}}$, we solve Eq.~\eqref{mbc2} using a root-finding method to detemine the possible values of $\alpha$ associated with stationary clouds. Only a discrete set of solutions, indexed by the non-negative integer $n$ that corresponds to the number of nodes in the cloud, exists.

We remark that the stationary configurations obtained by the aforementioned procedure lie at the threshold of superradiant instabilities and, as such, are expected to be marginally stable.

\section{Results}
\begin{figure}[ht]
\includegraphics[width=8.6cm]{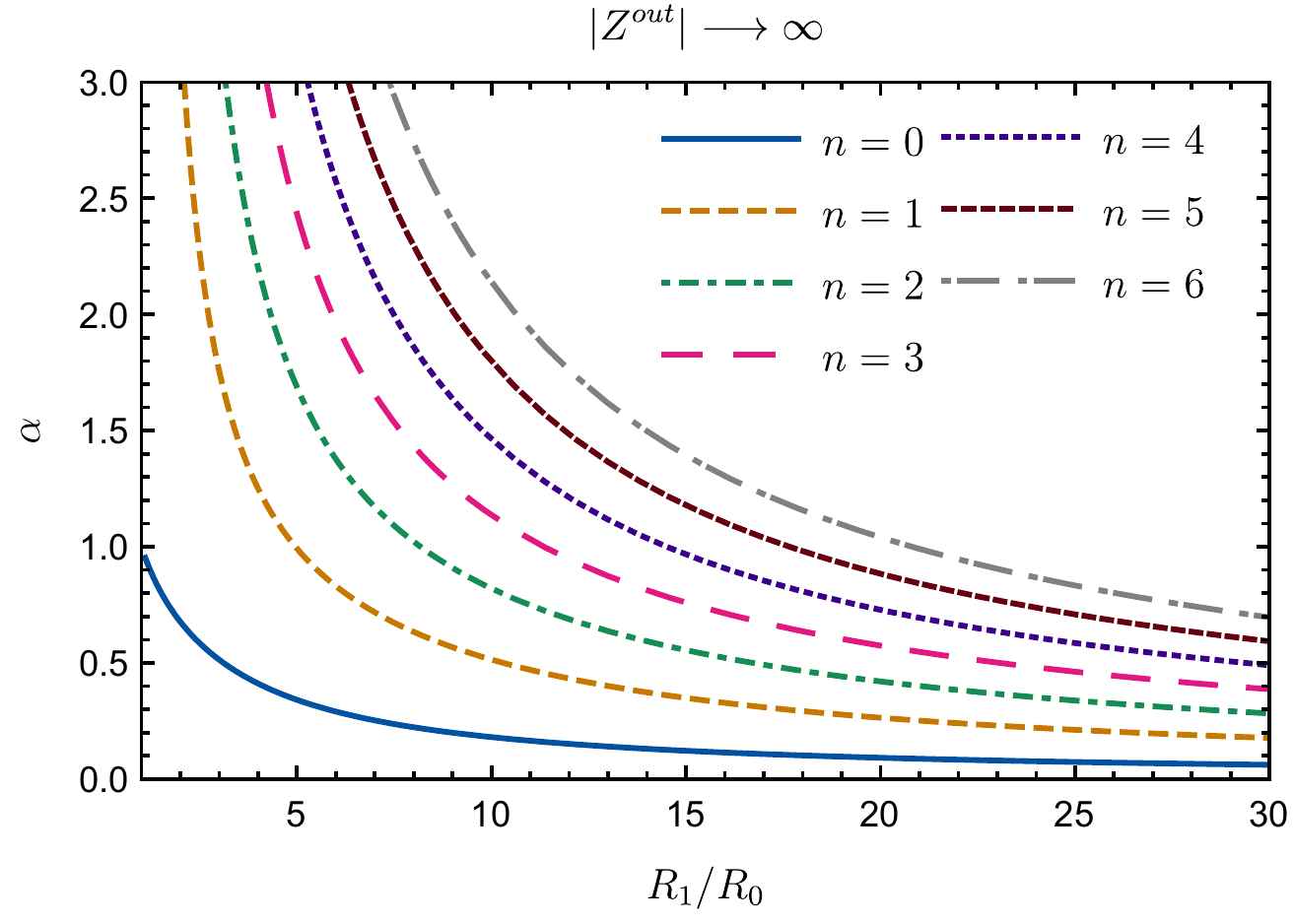}   \vspace{0.1in}\\
\includegraphics[width=8.6cm]{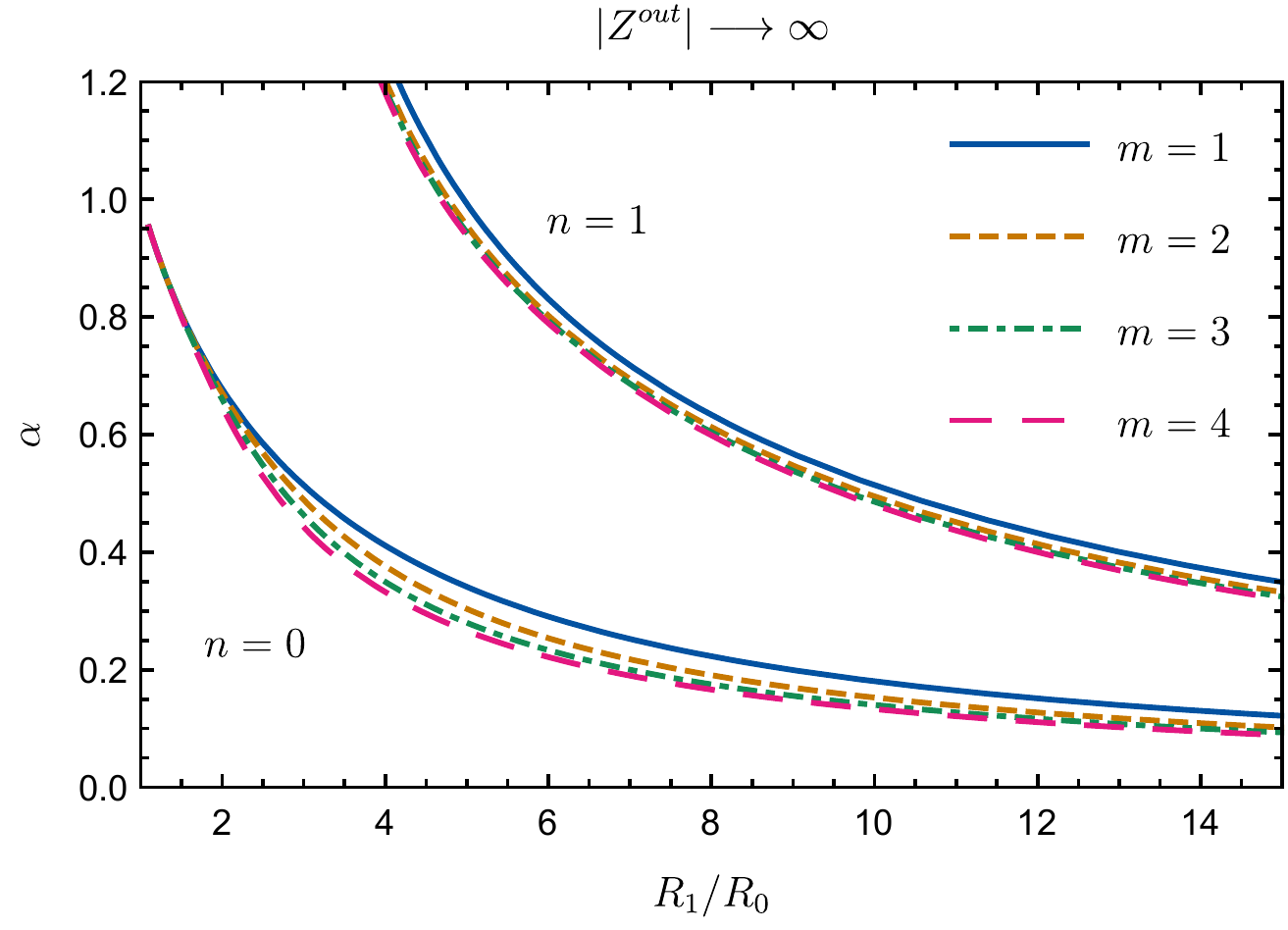}
\caption{Existence lines for stationary clouds considering the outer cylinder to be a hard wall. The top plot shows the first possible clouds (indexed by the node number $n$) for a fixed $m=1$ azimuthal number. The bottom plot exhibits the clouds with $n=0$ and $n=1$, for $m$ varying from $1$ to $4$.   
\label{eln}
}
\end{figure}

The first case we consider is that of a perfect reflector $|Z^{out}|\rightarrow  \infty$ (the hard wall limit). This corresponds to a Neumann boundary condition for $\psi$ at $r=R_1$, $cf.$ Eq.~\eqref{BCcyl0}. We show in Fig.~\ref{eln} (top panel) the associated clouds lying along one dimensional \textit{existence lines} in the two dimensional phase space formed by $\alpha$ and $R_1/R_0$. As a first trend, we observe (in the top panel) that clouds with a larger node number, $n$, for fixed $R_1/R_0$,  require a larger $\alpha$. That is, the more excited states require a larger angular velocity of the cylinder. We also observe that as $R_1/R_0\rightarrow \infty$, $\alpha \rightarrow 0$. This also seems natural, as we do not expect clouds to exist when the internal cylinder is static: the existence lines do not cross the abscissa axis. In the bottom panel of Fig.~\ref{eln} we see that as we increase $m$, for fixed $\alpha$ and $n$, the corresponding cloud occurs for smaller values of $R_1/R_0$. This is due to an equilibrium between the angular momentum of the cloud and of the internal cylinder, given by the condition $\omega/m=\Omega$.

\begin{figure}[ht]
\includegraphics[width=8.6cm]{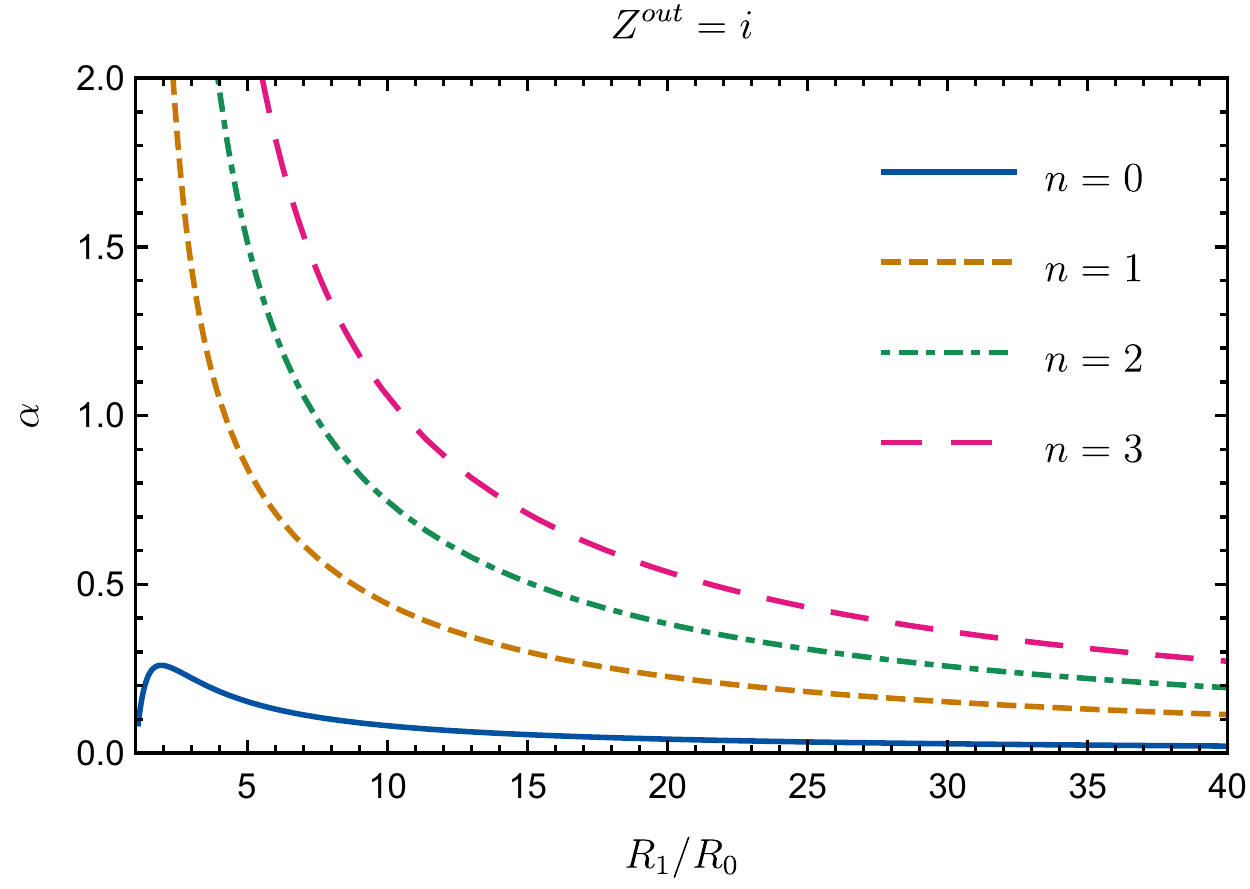}\vspace{0.1in} \\
\includegraphics[width=8.6cm]{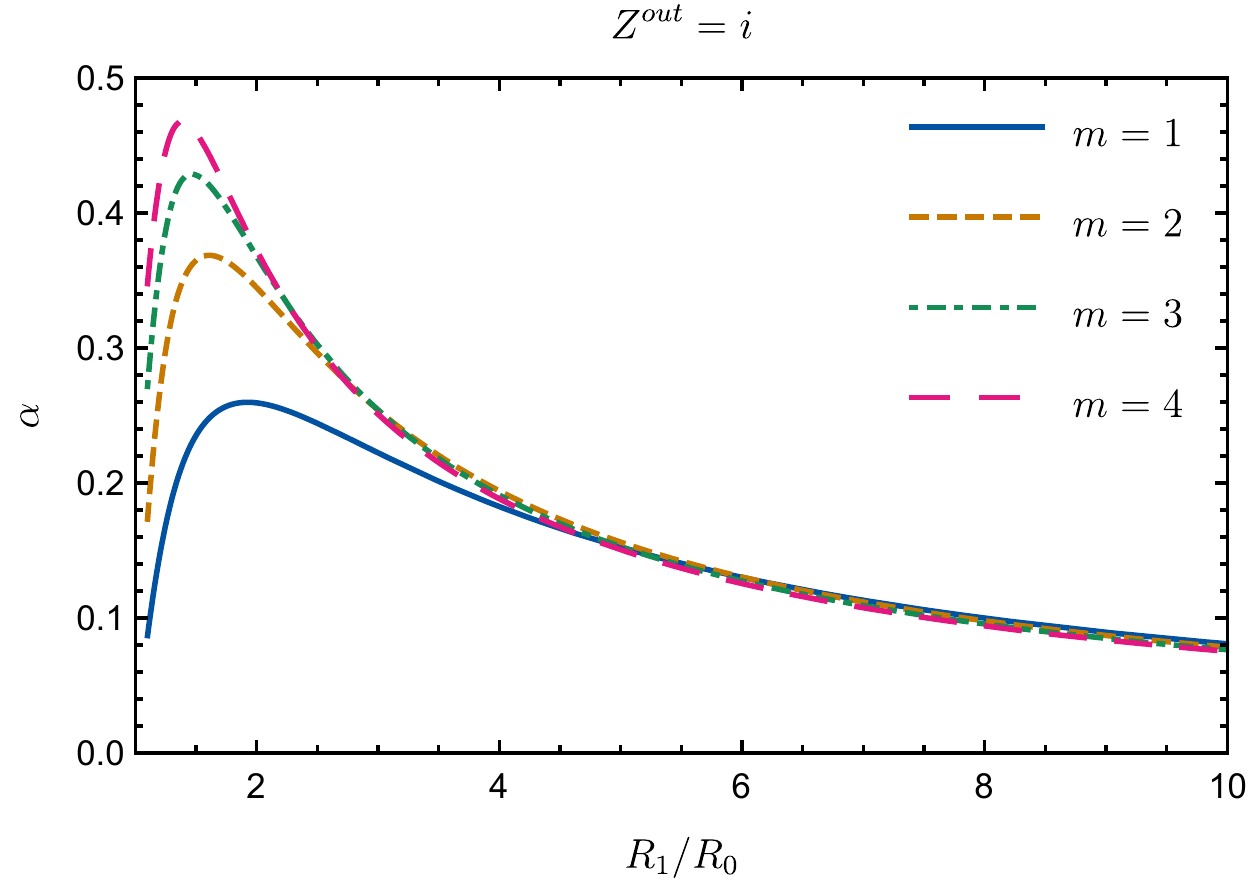}
\caption{Existence lines for stationary clouds when the outer cylinder is a compliance-dominated wall with $Z^{out}=i$. The top panel shows the $m=1$ cloud for $n=0,1,2,3$. The bottom panel exhibits the peculiar behavior of the $n=0$ cloud for $m=1,2,3,4$.}
\label{eli}
\end{figure}

The same detailed analysis presented above was also performed for other values of impedance. A  qualitatively novel, unexpected, feature was observed for impedances dominated by compliance rather than inertance. Using $Z^{out}=i$ as an example, we see in Fig.~\ref{eli} that the existence line for $n=0$ presents a maximum value of $\alpha$ at some value $R_1/R_0> 1$ (top panel). 
 In particular, for a fixed $\alpha$ below this maximum, fundamental ($n=0$) clouds are allowed at two different ratios $R_1/R_0$, while above it fundamental clouds are not allowed. Excited modes, on the other hand, did not exhibit a similar maximum in our analysis.  For the fundamental mode, the bottom panel of Fig.~\ref{eli} shows that, as we increase $m$, the maximum value of $\alpha$ increases and the value of $R_1/R_0$ at which the maximum occurs decreases. 
This maximum occurs when the two cylinders are positioned close to each other. Some insight on why the phase difference between incoming and outgoing waves gives rise to this maximum for a compliant, but not for an inert, impedance would be desirable. It is perhaps related to the fact that waves scattered by the wall can be confined to a thin layer near the wall and behave like a surface wave~\cite{Rienstra}.

\begin{figure}[ht]
\includegraphics[width=8.6cm]{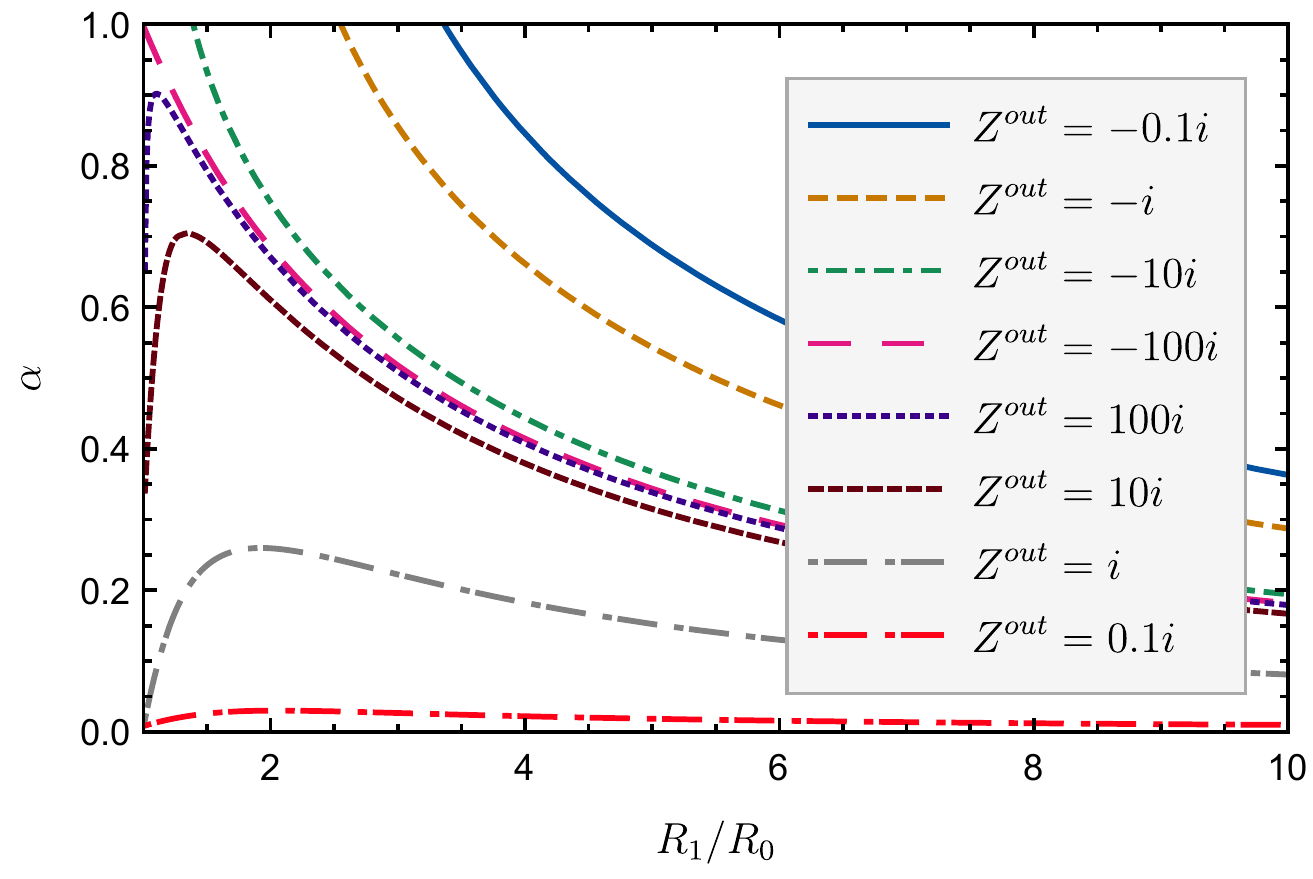}
\caption{Existence lines for different impedances of the outer cylinder, contrasting the cases of a hard wall, a soft-compliant wall, and a soft-inert wall. The cloud parameters are fixed as $m=1$ and $n=0$.}
\label{compz}
\end{figure}

The effect of a varying impedance $Z^{out}$ on the existence lines of the clouds is highlighted in Fig.~\ref{compz} for $n=0$ and $m=1$. We see that, for $\mathrm{Im}(Z^{out})<0$ and fixed $R_1/R_0$, as we increase $|Z^{out}|$, clouds occur for smaller values of $\alpha$. By contrast, for $\mathrm{Im}(Z^{out})>0$ and fixed $R_1/R_0$, clouds occur for larger values of $\alpha$ as $|Z^{out}|$ increases.  Note, in particular, that for $\mathrm{Im}(Z^{out})>0$, as we increase $|Z^{out}|$ the maximum of the existence line approaches $R_1/R_0=1$. When $\mathrm{Im}(Z^{out})\rightarrow \pm \infty$, the existence lines for both inertia-dominated and compliance-dominated impedances converge to one another, and to those exhibited in Fig.~\ref{eln}. This expected convergence is in agreement with the observed proximity between the $Z^{out}=\pm100i$ existence lines in Fig.~\ref{compz}.

\begin{figure}[ht]
\begin{picture}(100,100)
\put(0,50){\includegraphics[width=0.23\linewidth]{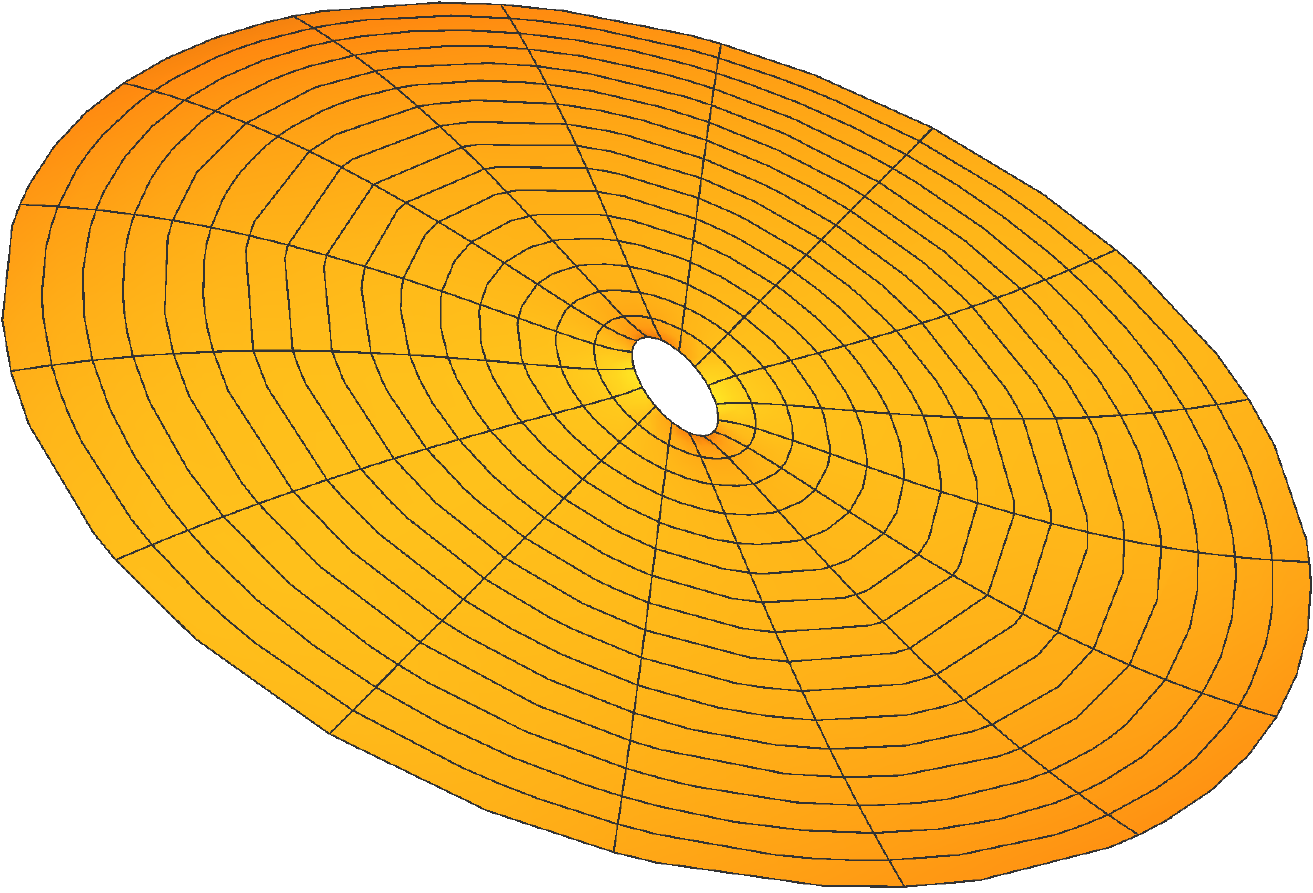}}
\put(20,100){$n=0$}
\put(90,50){\includegraphics[width=0.25\linewidth]{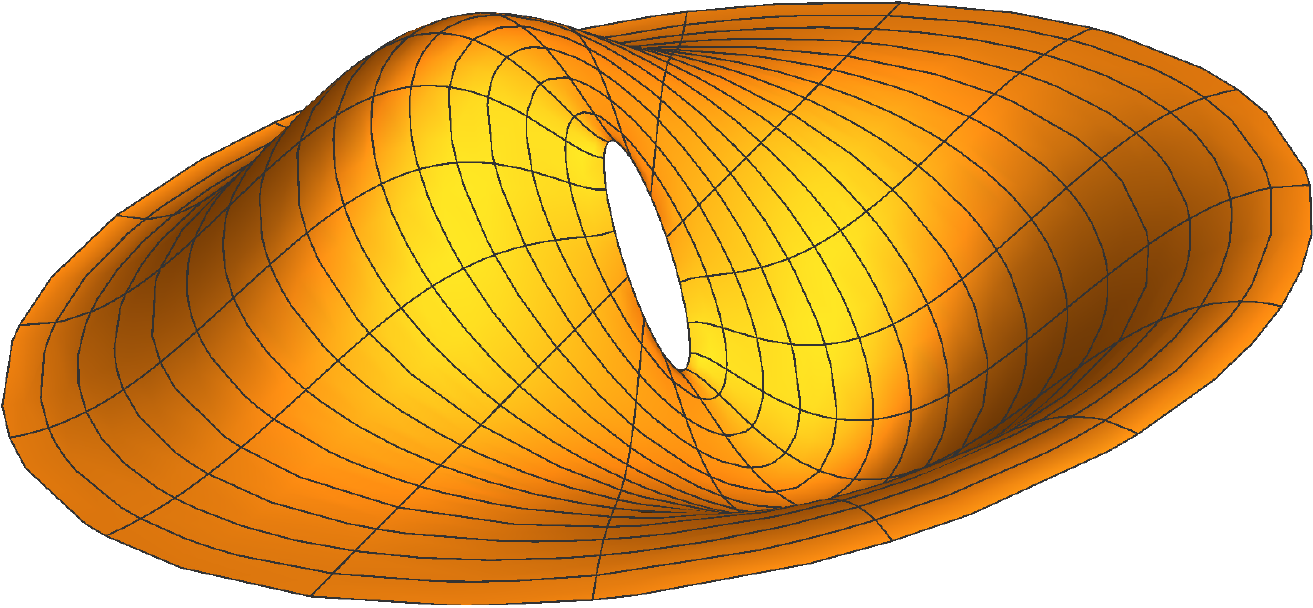}}
\put(110,100){$n=1$}
\put(180,50){\includegraphics[width=0.25\linewidth]{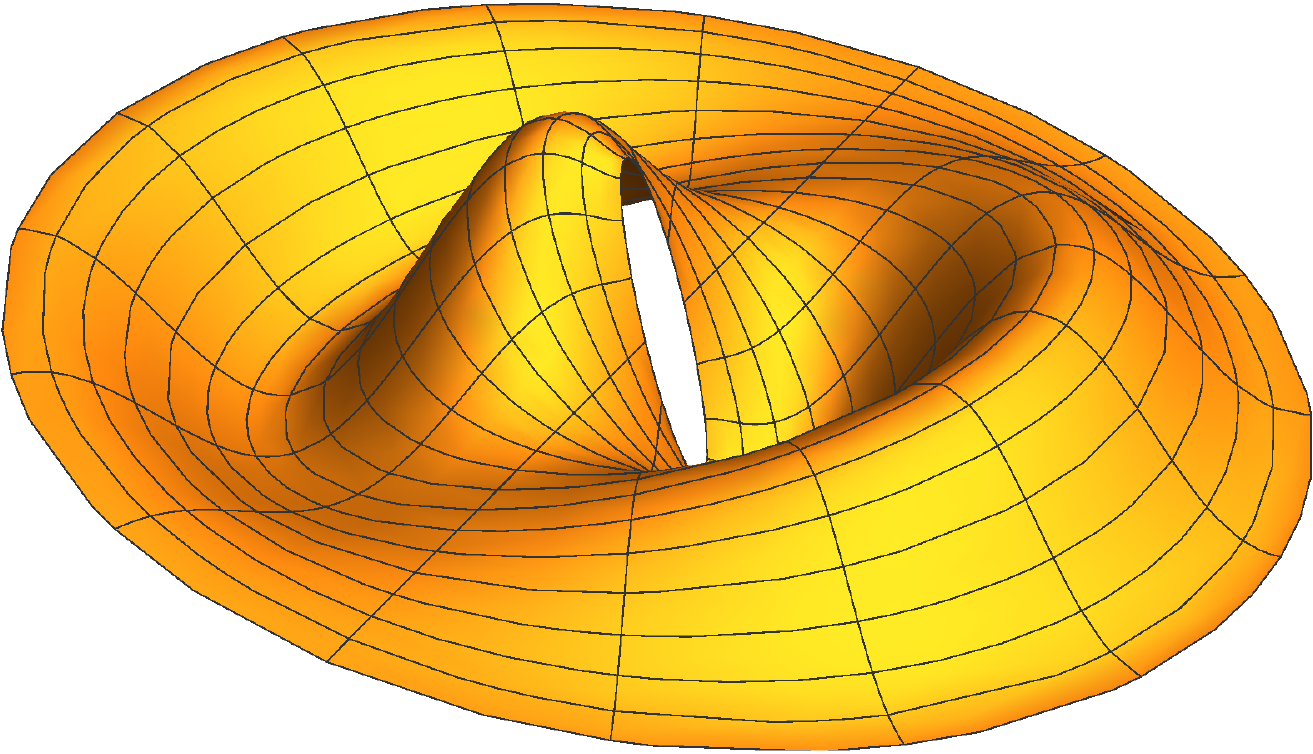}}
\put(200,100){$n=2$}
\put(0,0){\includegraphics[width=0.25\linewidth]{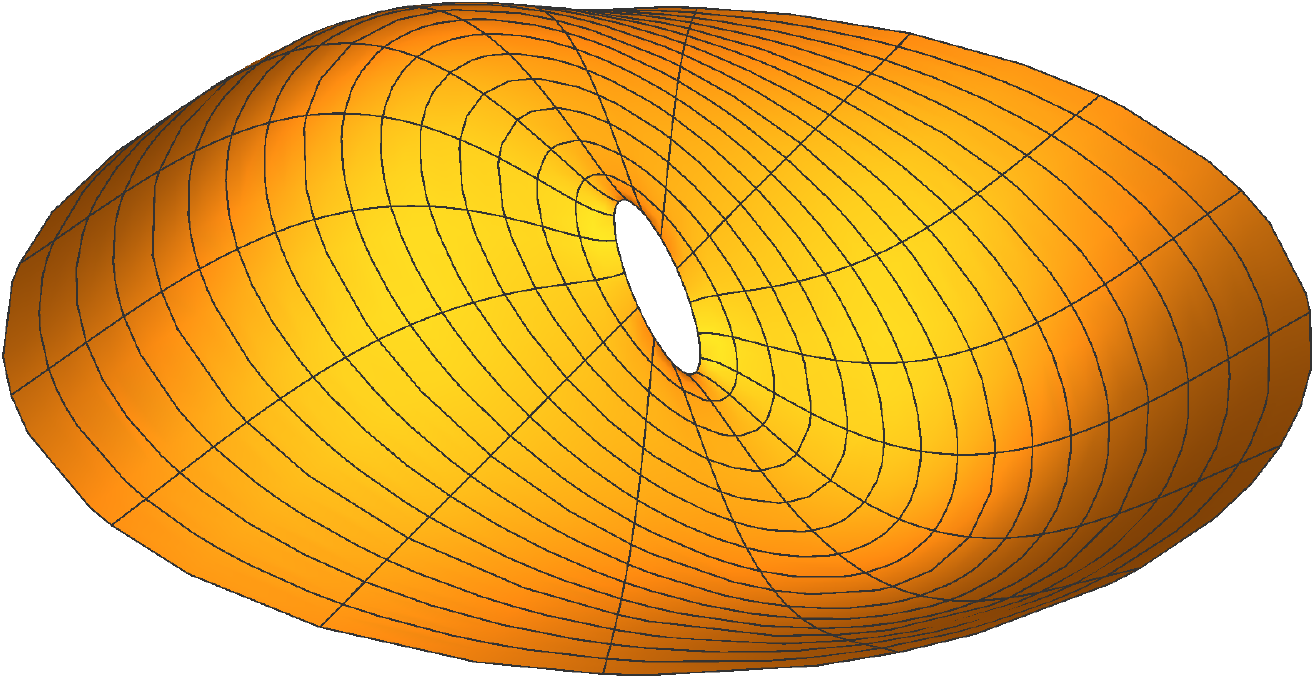}}
\put(90,0){\includegraphics[width=0.25\linewidth]{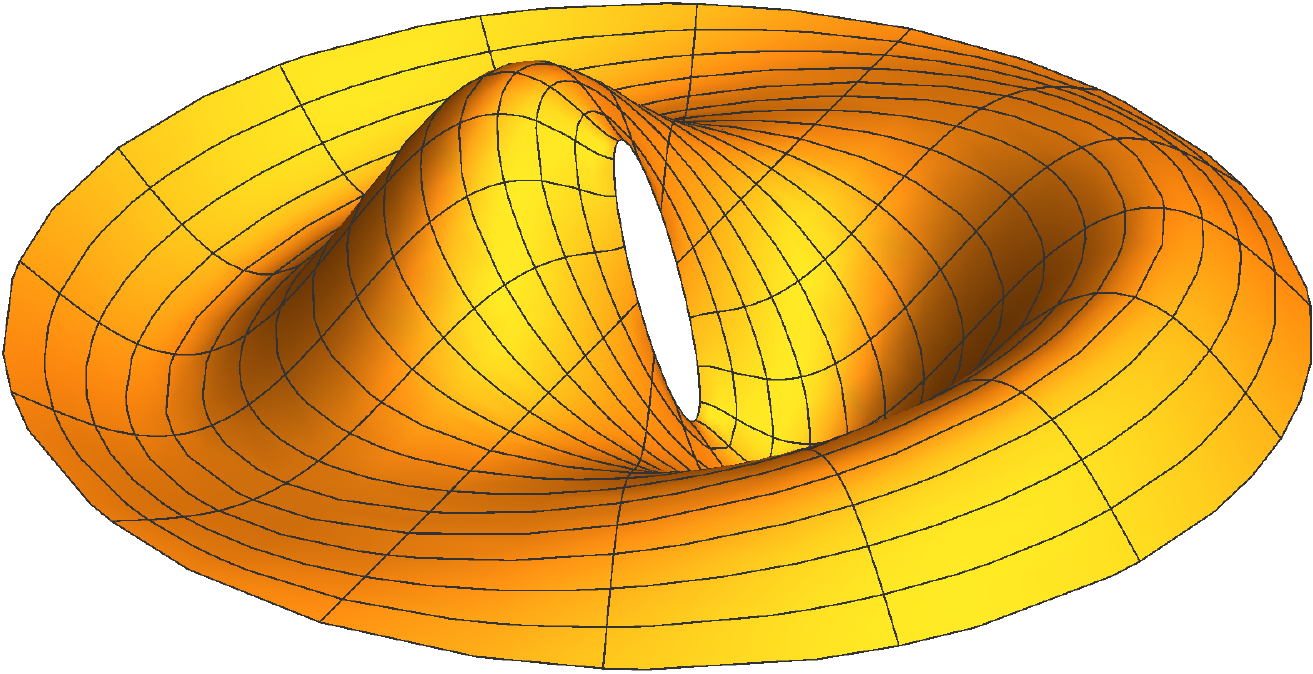}}
\put(180,0){\includegraphics[width=0.25\linewidth]{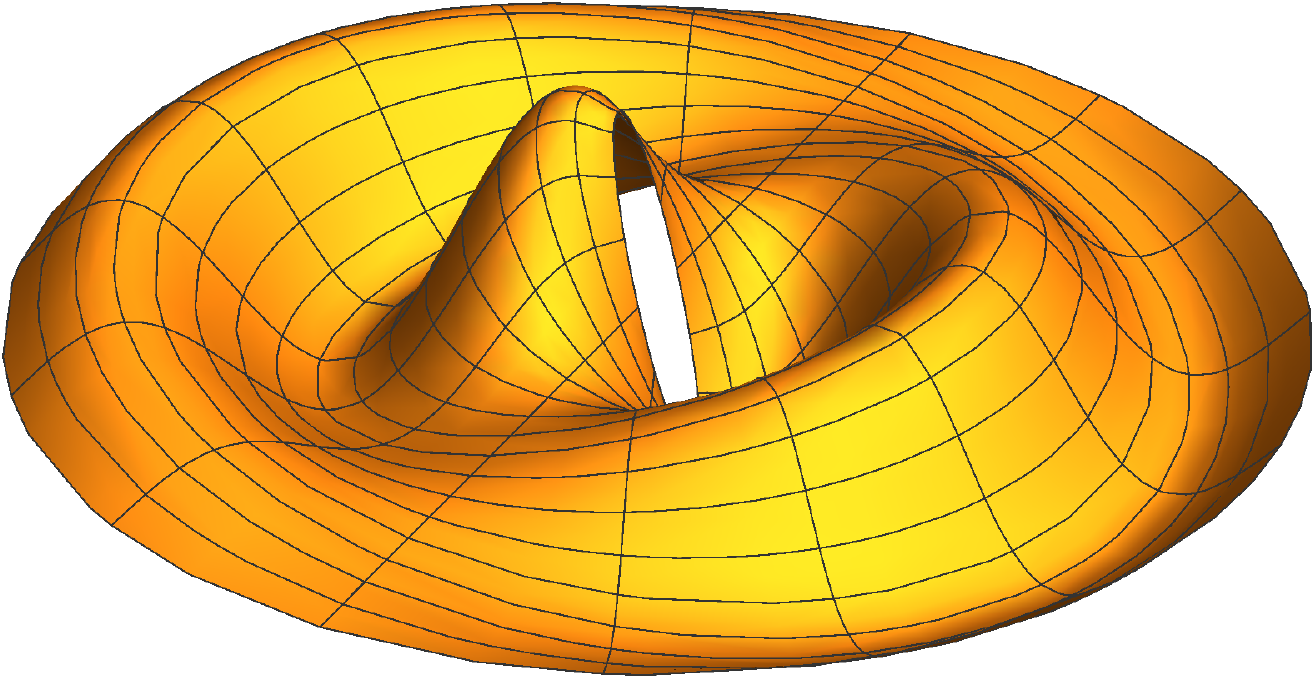}}
\end{picture}
\caption{Spatial distribution of the clouds with $m=1$ and $R_1/R_0=15$ in the $x, y$ plane. The upper plots were obtained considering the outer cylinder to be a hard wall, while the lower plots were obtained considering the outer cylinder to be a soft compliance-dominated wall.}
\label{tbl}
\end{figure}
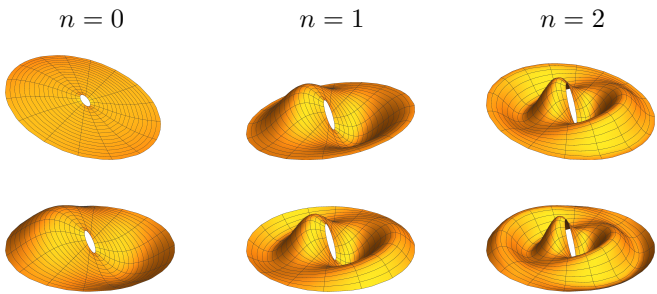

In Fig.~\ref{tbl} we exhibit the real part of the wave field $\psi$ at $t=0$
assuming, for the outer cylinder, either the hard wall limit (upper plots) or the soft-compliant wall limit (lower plots). We have fixed $m=1$, $R_1/R_0=15$ and, for simplicity $D_2=1$ in Eq.~\eqref{wavesol}. In particular, we observe how different values of $\alpha$ correspond to different node numbers $n$, for fixed ratios $R_1/R_0$, $cf.$~Fig.~\ref{eln}. The Neumann (Dirichlet) boundary condition associated with the hard (soft) wall means that
$\partial \psi/\partial r$ ($\psi$) vanishes at the external cylinder, as can be seen in the top (bottom) plots of Fig.~\ref{tbl}.

\section{Final Remarks}
The setup presented here can, in principle, be implemented in a laboratory aiming to detect such clouds experimentally. For instance, if shallow water surface waves are used, and $\alpha$ and $R_1/R_0$ are carefully chosen, we expect the surface of the water to settle to the configuration of a stationary cloud, as illustrated in Fig.~\ref{tbl}. One may worry, however, that a frequency fine tuning is needed to accomplish this and avoid superradiant instabilities. Fortunately, the time scale associated with these instabilities depends on the impedance of the inner cylinder. We have verified, by following the steps in \cite{cylinder}, that the harder the inner cylinder is, the slower the superradiant instabilities grow. If the inner cylinder is a hard wall, the instability time scale is arbitrarily large and a wave packet peaked around the synchronization frequency suffices to excite the desired cloud.

 As an example, assume a fluid laboratory setup with $R_0 = 0.05\, {\rm m}$, $R_1 = 0.75\, {\rm m}$, and $h = 0.025 {\rm m}$ (so that $c \sim 0.5 \, {\rm m }\, {\rm s}^{-1}$), which is consistent with the estimates in Ref.~\cite{cylinder}. Taking the outer cylinder to be a hard wall, we determine that the range of rotation speeds  needed to observe all seven clouds shown in Fig.~\ref{eln} (top panel) is $\Omega\sim 1 - 15 \, {\rm s}^{-1}$. Viscosity effects on wave propagation can be neglected unless the frequencies involved are extremely high, of order $10^6 \, {\rm s}^{-1}$~\cite{cylinder}, which is not the case  here. Diffusion of angular momentum from the rotating cylinder into the fluid can also be neglected as long as the boundary layer of rotating fluid that forms around the cylinder is sufficiently small compared to the wavelength of the clouds~\cite{cylinder}. Long wavelengths are also required by the shallow water wave limit that enforces the validity of Eq.~\eqref{eq:radial}. This is certainly the case for the parameters above, at least for the fundamental clouds. By tuning the rotation speed of the cylinder, different m modes of the fundamental cloud would be seen.  The observation of such configurations would be the first experimental detection of these bound states (i.e.~\textit{synchronized} stationary clouds) which are ubiquitous to rotating systems with dissipation.

Let us close by remarking that, as emphasized above, the stationary fluid configurations discussed herein rely on a synchronization condition between the frequency of the oscillating cloud, which is generically non-vanishing, and the rotation angular velocity of the inner cylinder. \textit{Static}, rather than stationary, equilibrium configurations in a fluid are also possible, with zero frequency, as discussed in~\cite{Hod:2014hda,Hod:2017lho}.

\section*{Acknowledgements}
The authors would like to thank Conselho Nacional de Desenvolvimento Cient\'ifico e Tecnol\'ogico (CNPq) and Coordena\c{c}\~ao de Aperfei\c{c}oamento de Pessoal de N\'ivel Superior (CAPES), from Brazil, for partial financial support.
This work has been supported by the FCT (Portugal) IF programme and  grant   PTDC/FIS-
OUT/28407/2017, the CIDMA (FCT) 
strategic project UID/MAT/04106/2013 and by  the  European  Union's  Horizon  2020  research  and  innovation  (RISE) programmes H2020-MSCA-RISE-2015 Grant No.~StronGrHEP-690904 and H2020-MSCA-RISE-2017 Grant No.~FunFiCO-777740. The authors would like to acknowledge networking support by the
COST Action CA16104. 
M.R. acknowledges partial support from the S\~ao Paulo Research Foundation (FAPESP), Grant No. 2013/09357-9.

\bibliography{clouds_cylinder_letter_revised_Oct}

\begin{thebibliography}{10}
\expandafter\ifx\csname url\endcsname\relax
  \def\url#1{\texttt{#1}}\fi
\expandafter\ifx\csname urlprefix\endcsname\relax\def\urlprefix{URL }\fi
\expandafter\ifx\csname href\endcsname\relax
  \def\href#1#2{#2} \def\path#1{#1}\fi

\bibitem{review1}
K.~D. Kokkotas, B.~G. Schmidt, {Quasinormal modes of stars and black holes},
  Living Rev. Rel. 2 (1999) 2.
\newblock \href {http://arxiv.org/abs/gr-qc/9909058}
  {\path{arXiv:gr-qc/9909058}}, \href {http://dx.doi.org/10.12942/lrr-1999-2}
  {\path{doi:10.12942/lrr-1999-2}}.

\bibitem{review2}
H.-P. Nollert, {TOPICAL REVIEW: Quasinormal modes: the characteristic `sound'
  of black holes and neutron stars}, Class.Quant.Grav. 16 (1999) R159--R216.
\newblock \href {http://dx.doi.org/10.1088/0264-9381/16/12/201}
  {\path{doi:10.1088/0264-9381/16/12/201}}.

\bibitem{Berti:2009kk}
E.~Berti, V.~Cardoso, A.~O. Starinets, {Quasinormal modes of black holes and
  black branes}, Class. Quant. Grav. 26 (2009) 163001.
\newblock \href {http://arxiv.org/abs/0905.2975} {\path{arXiv:0905.2975}},
  \href {http://dx.doi.org/10.1088/0264-9381/26/16/163001}
  {\path{doi:10.1088/0264-9381/26/16/163001}}.

\bibitem{Konoplya:2011qq}
R.~A. Konoplya, A.~Zhidenko, {Quasinormal modes of black holes: From
  astrophysics to string theory}, Rev. Mod. Phys. 83 (2011) 793--836.
\newblock \href {http://arxiv.org/abs/1102.4014} {\path{arXiv:1102.4014}},
  \href {http://dx.doi.org/10.1103/RevModPhys.83.793}
  {\path{doi:10.1103/RevModPhys.83.793}}.

\bibitem{zeldovich1}
Y.~B. {Zel'Dovich}, {Generation of Waves by a Rotating Body}, JETP Lett. 14
  (1971) 180.

\bibitem{bek_1}
J.~D. Bekenstein, M.~Schiffer, {The Many faces of superradiance}, Phys. Rev.
  D58 (1998) 064014.
\newblock \href {http://arxiv.org/abs/gr-qc/9803033}
  {\path{arXiv:gr-qc/9803033}}, \href
  {http://dx.doi.org/10.1103/PhysRevD.58.064014}
  {\path{doi:10.1103/PhysRevD.58.064014}}.

\bibitem{Richartz:2009mi}
M.~{Richartz}, S.~{Weinfurtner}, A.~J. {Penner}, W.~G. {Unruh}, {Generalized
  superradiant scattering}, Phys.Rev. D80~(12) (2009) 124016.
\newblock \href {http://arxiv.org/abs/0909.2317} {\path{arXiv:0909.2317}},
  \href {http://dx.doi.org/10.1103/PhysRevD.80.124016}
  {\path{doi:10.1103/PhysRevD.80.124016}}.

\bibitem{Brito:2015oca}
R.~Brito, V.~Cardoso, P.~Pani, {Superradiance}, Lect. Notes Phys. 906 (2015)
  pp.1--237.
\newblock \href {http://arxiv.org/abs/1501.06570} {\path{arXiv:1501.06570}},
  \href {http://dx.doi.org/10.1007/978-3-319-19000-6}
  {\path{doi:10.1007/978-3-319-19000-6}}.

\bibitem{Hod:2012px}
S.~Hod, {Stationary Scalar Clouds Around Rotating Black Holes}, Phys.Rev. D86
  (2012) 104026.
\newblock \href {http://arxiv.org/abs/1211.3202} {\path{arXiv:1211.3202}},
  \href {http://dx.doi.org/10.1103/PhysRevD.86.129902,
  10.1103/PhysRevD.86.104026} {\path{doi:10.1103/PhysRevD.86.129902,
  10.1103/PhysRevD.86.104026}}.

\bibitem{Hod:2014baa}
S.~Hod, {Kerr-Newman black holes with stationary charged scalar clouds},
  Phys.Rev. D90 (2014) 024051.
\newblock \href {http://arxiv.org/abs/1406.1179} {\path{arXiv:1406.1179}}.

\bibitem{Hod:2013zza}
S.~Hod, {Stationary resonances of rapidly-rotating Kerr black holes},
  Eur.Phys.J. C73 (2013) 2378.
\newblock \href {http://arxiv.org/abs/1311.5298} {\path{arXiv:1311.5298}},
  \href {http://dx.doi.org/10.1140/epjc/s10052-013-2378-x}
  {\path{doi:10.1140/epjc/s10052-013-2378-x}}.

\bibitem{benone2014kerr}
C.~L. Benone, L.~C. Crispino, C.~Herdeiro, E.~Radu, Kerr-newman scalar clouds,
  Physical Review D 90~(10) (2014) 104024.

\bibitem{Benone:2014nla}
C.~L. Benone, L.~C.~B. Crispino, C.~Herdeiro, E.~Radu, {Acoustic clouds:
  standing sound waves around a black hole analogue}, Phys. Rev. D91~(10)
  (2015) 104038.
\newblock \href {http://arxiv.org/abs/1412.7278} {\path{arXiv:1412.7278}},
  \href {http://dx.doi.org/10.1103/PhysRevD.91.104038}
  {\path{doi:10.1103/PhysRevD.91.104038}}.

\bibitem{Hod:2014pza}
S.~Hod, {Onset of superradiant instabilities in the composed
  Kerr-black-hole-mirror bomb}, Phys. Lett. B736 (2014) 398--402.
\newblock \href {http://arxiv.org/abs/1412.6108} {\path{arXiv:1412.6108}},
  \href {http://dx.doi.org/10.1016/j.physletb.2014.07.049}
  {\path{doi:10.1016/j.physletb.2014.07.049}}.

\bibitem{Hod:2014sha}
S.~Hod, {Rotating black holes can have short bristles}, Phys.Lett. B739 (2014)
  196.
\newblock \href {http://arxiv.org/abs/1411.2609} {\path{arXiv:1411.2609}}.

\bibitem{Herdeiro:2014pka}
C.~Herdeiro, E.~Radu, H.~Runarsson, {Non-linear $Q$-clouds around Kerr black
  holes}, Phys.Lett. B739 (2014) 302--307.
\newblock \href {http://arxiv.org/abs/1409.2877} {\path{arXiv:1409.2877}},
  \href {http://dx.doi.org/10.1016/j.physletb.2014.11.005}
  {\path{doi:10.1016/j.physletb.2014.11.005}}.

\bibitem{Wilson-Gerow:2015esa}
J.~Wilson-Gerow, A.~Ritz, {Black hole energy extraction via a stationary scalar
  analog of the Blandford-Znajek mechanism}, Phys. Rev. D93~(4) (2016) 044043.
\newblock \href {http://arxiv.org/abs/1509.06681} {\path{arXiv:1509.06681}},
  \href {http://dx.doi.org/10.1103/PhysRevD.93.044043}
  {\path{doi:10.1103/PhysRevD.93.044043}}.

\bibitem{Wang:2015fgp}
M.~Wang, C.~Herdeiro, {Maxwell perturbations on Kerr-anti-de Sitter black
  holes: Quasinormal modes, superradiant instabilities, and vector clouds},
  Phys. Rev. D93~(6) (2016) 064066.
\newblock \href {http://arxiv.org/abs/1512.02262} {\path{arXiv:1512.02262}},
  \href {http://dx.doi.org/10.1103/PhysRevD.93.064066}
  {\path{doi:10.1103/PhysRevD.93.064066}}.

\bibitem{0264-9381-33-15-154001}
C.~Herdeiro, E.~Radu, H.~Runarsson, {Kerr black holes with Proca hair}, Class.
  Quant. Grav. 33~(15) (2016) 154001.
\newblock \href {http://arxiv.org/abs/1603.02687} {\path{arXiv:1603.02687}},
  \href {http://dx.doi.org/10.1088/0264-9381/33/15/154001}
  {\path{doi:10.1088/0264-9381/33/15/154001}}.

\bibitem{Bernard:2016wqo}
C.~Bernard, {Stationary charged scalar clouds around black holes in string
  theory}, Phys. Rev. D94~(8) (2016) 085007.
\newblock \href {http://arxiv.org/abs/1608.05974} {\path{arXiv:1608.05974}},
  \href {http://dx.doi.org/10.1103/PhysRevD.94.085007}
  {\path{doi:10.1103/PhysRevD.94.085007}}.

\bibitem{Sakalli:2016xoa}
I.~Sakalli, G.~Tokgoz, {Stationary Scalar Clouds Around Maximally Rotating
  Linear Dilaton Black Holes}, Class. Quant. Grav. 34~(12) (2017) 125007.
\newblock \href {http://arxiv.org/abs/1610.09329} {\path{arXiv:1610.09329}},
  \href {http://dx.doi.org/10.1088/1361-6382/aa6858}
  {\path{doi:10.1088/1361-6382/aa6858}}.

\bibitem{Hod:2016lgi}
S.~Hod, {Spinning Kerr black holes with stationary massive scalar clouds: The
  large-coupling regime}, JHEP 01 (2017) 030.
\newblock \href {http://arxiv.org/abs/1612.00014} {\path{arXiv:1612.00014}},
  \href {http://dx.doi.org/10.1007/JHEP01(2017)030}
  {\path{doi:10.1007/JHEP01(2017)030}}.

\bibitem{Hod:2016yxg}
S.~Hod, {The large-mass limit of cloudy black holes}, Class. Quant. Grav.
  32~(13) (2015) 134002.
\newblock \href {http://arxiv.org/abs/1607.00003} {\path{arXiv:1607.00003}},
  \href {http://dx.doi.org/10.1088/0264-9381/32/13/134002}
  {\path{doi:10.1088/0264-9381/32/13/134002}}.

\bibitem{Herdeiro:2017oyt}
C.~Herdeiro, J.~Kunz, E.~Radu, B.~Subagyo, {Probing the universality of
  synchronised hair around rotating black holes with Q-clouds}, Phys. Lett.
  B779 (2018) 151--159.
\newblock \href {http://arxiv.org/abs/1712.04286} {\path{arXiv:1712.04286}},
  \href {http://dx.doi.org/10.1016/j.physletb.2018.01.083}
  {\path{doi:10.1016/j.physletb.2018.01.083}}.

\bibitem{Ferreira:2017cta}
H.~R.~C. Ferreira, C.~A.~R. Herdeiro, {Stationary scalar clouds around a BTZ
  black hole}, Phys. Lett. B773 (2017) 129--134.
\newblock \href {http://arxiv.org/abs/1707.08133} {\path{arXiv:1707.08133}},
  \href {http://dx.doi.org/10.1016/j.physletb.2017.08.017}
  {\path{doi:10.1016/j.physletb.2017.08.017}}.

\bibitem{Herdeiro:2014goa}
C.~A.~R. Herdeiro, E.~Radu, {Kerr black holes with scalar hair}, Phys.Rev.Lett.
  112 (2014) 221101.
\newblock \href {http://arxiv.org/abs/1403.2757} {\path{arXiv:1403.2757}},
  \href {http://dx.doi.org/10.1103/PhysRevLett.112.221101}
  {\path{doi:10.1103/PhysRevLett.112.221101}}.

\bibitem{Herdeiro:2015gia}
C.~Herdeiro, E.~Radu, {Construction and physical properties of Kerr black holes
  with scalar hair}, Class. Quant. Grav. 32~(14) (2015) 144001.
\newblock \href {http://arxiv.org/abs/1501.04319} {\path{arXiv:1501.04319}},
  \href {http://dx.doi.org/10.1088/0264-9381/32/14/144001}
  {\path{doi:10.1088/0264-9381/32/14/144001}}.

\bibitem{Chodosh:2015oma}
O.~Chodosh, Y.~Shlapentokh-Rothman, {Time-Periodic Einstein-Klein-Gordon
  Bifurcations of Kerr}, Commun. Math. Phys. 356~(3) (2017) 1155--1250.
\newblock \href {http://arxiv.org/abs/1510.08025} {\path{arXiv:1510.08025}},
  \href {http://dx.doi.org/10.1007/s00220-017-2998-3}
  {\path{doi:10.1007/s00220-017-2998-3}}.

\bibitem{East:2017ovw}
W.~E. East, F.~Pretorius, {Superradiant Instability and Backreaction of Massive
  Vector Fields around Kerr Black Holes}, Phys. Rev. Lett. 119~(4) (2017)
  041101.
\newblock \href {http://arxiv.org/abs/1704.04791} {\path{arXiv:1704.04791}},
  \href {http://dx.doi.org/10.1103/PhysRevLett.119.041101}
  {\path{doi:10.1103/PhysRevLett.119.041101}}.

\bibitem{Herdeiro:2017phl}
C.~A.~R. Herdeiro, E.~Radu, {Dynamical Formation of Kerr Black Holes with
  Synchronized Hair: An Analytic Model}, Phys. Rev. Lett. 119~(26) (2017)
  261101.
\newblock \href {http://arxiv.org/abs/1706.06597} {\path{arXiv:1706.06597}},
  \href {http://dx.doi.org/10.1103/PhysRevLett.119.261101}
  {\path{doi:10.1103/PhysRevLett.119.261101}}.

\bibitem{Ganchev:2017uuo}
B.~Ganchev, J.~E. Santos, {Scalar Hairy Black Holes in Four Dimensions are
  Unstable}, Phys. Rev. Lett. 120~(17) (2018) 171101.
\newblock \href {http://arxiv.org/abs/1711.08464} {\path{arXiv:1711.08464}},
  \href {http://dx.doi.org/10.1103/PhysRevLett.120.171101}
  {\path{doi:10.1103/PhysRevLett.120.171101}}.

\bibitem{Degollado:2018ypf}
J.~C. Degollado, C.~A.~R. Herdeiro, E.~Radu, {Effective stability against
  superradiance of Kerr black holes with synchronised hair}, Phys. Lett. B781
  (2018) 651--655.
\newblock \href {http://arxiv.org/abs/1802.07266} {\path{arXiv:1802.07266}},
  \href {http://dx.doi.org/10.1016/j.physletb.2018.04.052}
  {\path{doi:10.1016/j.physletb.2018.04.052}}.

\bibitem{Herdeiro:2015waa}
C.~A.~R. Herdeiro, E.~Radu, {Asymptotically flat black holes with scalar hair:
  a review}, Int. J. Mod. Phys. D24~(09) (2015) 1542014.
\newblock \href {http://arxiv.org/abs/1504.08209} {\path{arXiv:1504.08209}},
  \href {http://dx.doi.org/10.1142/S0218271815420146}
  {\path{doi:10.1142/S0218271815420146}}.

\bibitem{Unruh:1980cg}
W.~Unruh, {Experimental black hole evaporation}, Phys.Rev.Lett. 46 (1981)
  1351--1353.
\newblock \href {http://dx.doi.org/10.1103/PhysRevLett.46.1351}
  {\path{doi:10.1103/PhysRevLett.46.1351}}.

\bibitem{Barcelo:2005fc}
C.~Barcelo, S.~Liberati, M.~Visser, {Analogue gravity}, Living Rev.Rel. 8
  (2005) 12.
\newblock \href {http://arxiv.org/abs/gr-qc/0505065}
  {\path{arXiv:gr-qc/0505065}}.

\bibitem{PhysRevLett.121.061101}
S.~Patrick, A.~Coutant, M.~Richartz, S.~Weinfurtner, Black hole quasibound
  states from a draining bathtub vortex flow, Phys. Rev. Lett. 121 (2018)
  061101.
\newblock \href {http://dx.doi.org/10.1103/PhysRevLett.121.061101}
  {\path{doi:10.1103/PhysRevLett.121.061101}}.

\bibitem{Richartz:2013hza}
M.~Richartz, A.~Saa, {Superradiance without event horizons in General
  Relativity}, Phys. Rev. D88 (2013) 044008.
\newblock \href {http://arxiv.org/abs/1306.3137} {\path{arXiv:1306.3137}},
  \href {http://dx.doi.org/10.1103/PhysRevD.88.044008}
  {\path{doi:10.1103/PhysRevD.88.044008}}.

\bibitem{Oliveira:2014oja}
L.~A. Oliveira, V.~Cardoso, L.~C.~B. Crispino, {Ergoregion instability: The
  hydrodynamic vortex}, Phys. Rev. D89~(12) (2014) 124008.
\newblock \href {http://arxiv.org/abs/1405.4038} {\path{arXiv:1405.4038}},
  \href {http://dx.doi.org/10.1103/PhysRevD.89.124008}
  {\path{doi:10.1103/PhysRevD.89.124008}}.

\bibitem{Oliveira:2018ckz}
L.~A. Oliveira, L.~J. Garay, L.~C.~B. Crispino, {Ergoregion instability of a
  rotating quantum system}, Phys. Rev. D97~(12) (2018) 124063.
\newblock \href {http://dx.doi.org/10.1103/PhysRevD.97.124063}
  {\path{doi:10.1103/PhysRevD.97.124063}}.

\bibitem{zeldovich2}
Y.~B. {Zel'Dovich}, {Amplification of Cylindrical Electromagnetic Waves
  Reflected from a Rotating Body}, Sov. Phys. JETP 35 (1972) 1085.

\bibitem{cylinder}
V.~Cardoso, A.~Coutant, M.~Richartz, S.~Weinfurtner, {Detecting Rotational
  Superradiance in Fluid Laboratories}, Phys. Rev. Lett. 117~(27) (2016)
  271101.
\newblock \href {http://arxiv.org/abs/1607.01378} {\path{arXiv:1607.01378}},
  \href {http://dx.doi.org/10.1103/PhysRevLett.117.271101}
  {\path{doi:10.1103/PhysRevLett.117.271101}}.

\bibitem{stim_HR}
S.~Weinfurtner, E.~W. Tedford, M.~C.~J. Penrice, W.~G. Unruh, G.~A. Lawrence,
  {Measurement of stimulated Hawking emission in an analogue system}, Phys.
  Rev. Lett. 106 (2011) 021302.
\newblock \href {http://arxiv.org/abs/1008.1911} {\path{arXiv:1008.1911}},
  \href {http://dx.doi.org/10.1103/PhysRevLett.106.021302}
  {\path{doi:10.1103/PhysRevLett.106.021302}}.

\bibitem{germain_watertank}
L.~P. Euv{\'e}, F.~Michel, R.~Parentani, T.~G. Philbin, G.~Rousseaux,
  {Observation of noise correlated by the Hawking effect in a water tank},
  Phys. Rev. Lett. 117~(12) (2016) 121301.
\newblock \href {http://arxiv.org/abs/1511.08145} {\path{arXiv:1511.08145}},
  \href {http://dx.doi.org/10.1103/PhysRevLett.117.121301}
  {\path{doi:10.1103/PhysRevLett.117.121301}}.

\bibitem{steinhauer}
J.~Steinhauer, {Observation of quantum Hawking radiation and its entanglement
  in an analogue black hole}, Nature Phys. 12 (2016) 959.
\newblock \href {http://arxiv.org/abs/1510.00621} {\path{arXiv:1510.00621}},
  \href {http://dx.doi.org/10.1038/nphys3863} {\path{doi:10.1038/nphys3863}}.

\bibitem{SR_obsvn}
T.~Torres, S.~Patrick, A.~Coutant, M.~Richartz, E.~W. Tedford, S.~Weinfurtner,
  {Observation of superradiance in a vortex flow}, Nature Phys. 13 (2017)
  833--836.
\newblock \href {http://arxiv.org/abs/1612.06180} {\path{arXiv:1612.06180}},
  \href {http://dx.doi.org/10.1038/nphys4151} {\path{doi:10.1038/nphys4151}}.

\bibitem{Rienstra}
S.~W. Rienstra, A.~Hirschberg, {An Introduction to
  Acoustics}\noindent\url{http://www.win.tue.nl/~sjoerdr/papers/boek.pdf}.

\bibitem{ginsberg}
J.~Ginsberg,
  \href{https://books.google.com.br/books?id=6JQ4DwAAQBAJ}{Acoustics-A Textbook
  for Engineers and Physicists: Volume I: Fundamentals}, Springer International
  Publishing, 2017.
\newline\urlprefix\url{https://books.google.com.br/books?id=6JQ4DwAAQBAJ}

\bibitem{kennelly}
A.~E. Kennelly, K.~Kurokawa,
  \href{http://www.jstor.org/stable/20025828}{Acoustic impedance and its
  measurement}, Proceedings of the American Academy of Arts and Sciences 56~(1)
  (1921) 3--42.
\newline\urlprefix\url{http://www.jstor.org/stable/20025828}

\bibitem{Hod:2014hda}
S.~Hod, {Onset of superradiant instabilities in the hydrodynamic vortex model},
  Phys.Rev. D90 (2014) 027501.
\newblock \href {http://arxiv.org/abs/1405.7702} {\path{arXiv:1405.7702}}.

\bibitem{Hod:2017lho}
S.~Hod, {Marginally stable resonant modes of the polytropic hydrodynamic
  vortex}, Phys. Lett. B774 (2017) 368.
\newblock \href {http://arxiv.org/abs/1711.02105} {\path{arXiv:1711.02105}},
  \href {http://dx.doi.org/10.1016/j.physletb.2017.09.087}
  {\path{doi:10.1016/j.physletb.2017.09.087}}.

\end{thebibliography}

\end{document}